 \documentclass[12pt,a4paper]{article}

\usepackage{setspace}
\usepackage{authblk}

\usepackage{float}
\restylefloat{table}

\usepackage{graphicx}	 			
\usepackage{amsmath}
\usepackage{amssymb}
\usepackage{mathrsfs}
\usepackage{amsthm}
\usepackage{ulem}
\usepackage{graphicx}	

\usepackage{cancel}

\usepackage{natbib}
\usepackage{epsfig,amstext,amsfonts,color}

\usepackage{latexsym}

\usepackage{url}

\newcommand{\E}{\operatorname{E}}

\newcommand{\bone}{\mathbf{1}}
\newcommand{\bzero}{\mathbf{0}}

\newcommand{\bbeta}{\mathbf{\beta}}
\newcommand{\btheta}{\mathbf{\theta}}

\newcommand{\bpi}{\mathbf{\pi}}

\newcommand{\bz}{\mathbf{z}}
\newcommand{\bx}{\mathbf{x}}
\newcommand{\by}{\mathbf{y}}

\newcommand{\bt}{\mathbf{t}}
\newcommand{\bu}{\mathbf{u}}

\newcommand{\bb}{\mathbf{b}}

\newcommand{\bd}{\mathbf{d}}
\newcommand{\bH}{\mathbf{H}}

\newcommand{\bR}{\mathbf{R}}

\newcommand{\bA}{\mathbf{A}}
\newcommand{\bG}{\mathbf{G}}
\newcommand{\bD}{\mathbf{D}}
\newcommand{\bS}{\mathbf{S}}

\newcommand{\bW}{\mathbf{W}}
\newcommand{\bB}{\mathbf{B}}

\renewcommand{\mathbf}{\boldsymbol}

\title{Robust score matching for compositional data}

\author[1,*]{Janice L. Scealy}
\author[1]{Kassel L. Hingee}
\author[2]{John T. Kent}
\author[1]{Andrew T. A. Wood}
\affil[1]{Research School of Finance, Actuarial Studies and Statistics, Australian National University, Canberra ACT 2601, Australia}
\affil[2]{Department of Statistics, University of Leeds, Leeds LS2 9JT, UK.}
\affil[*]{Corresponding author's email: janice.scealy@anu.edu.au.~~~~~~~~~~~~~~~~~~~~~~~~~~~~~~~~}

\date{}

\makeatletter
\renewcommand\@biblabel[1]{}
\makeatother

\begin{document}

\maketitle

\begin{abstract}
The restricted polynomially-tilted pairwise interaction (RPPI)
  distribution gives a flexible model for compositional data.  It is
  particularly well-suited to situations where some of the marginal
  distributions of the components of a composition are concentrated
  near zero, possibly with right skewness.  This article develops a
  method of tractable robust estimation for the model by combining two
  ideas.  The first idea is to use score matching estimation after an
  additive log-ratio transformation. The resulting estimator is automatically insensitive to zeros in the data compositions. The second idea is to
  incorporate suitable weights in the estimating equations.  The
  resulting estimator is additionally resistant to outliers.  These properties are confirmed in simulation studies where we further also demonstrate that our new outlier-robust estimator is efficient in high concentration settings, even in the case when there is no model contamination.  An example is given using microbiome data.
  A user-friendly R package accompanies the article.

\vspace{3.mm}
\end{abstract}	

\section{Introduction}
The polynomially-tilted pairwise interaction (PPI) model for
  compositional data was introduced by Scealy and Wood (2022).  It is
  a flexible model for compositional data because it can model high
  levels of right skewness in the marginal distributions of the
  components of a composition, and it can capture a wide range of
  correlation patterns.  Empirical investigations in Scealy and Wood
  (2022) showed that this distribution can successfully describe the
  behaviour of real data in many settings.  They illustrated its
  effectiveness on a set of microbiome data.

Most articles in the literature analysing microbiome data,  including for example Cao et al. (2019), He et al. (2021), Mishra and Muller (2022) and Liang et al. (2022), assume that zeros are outliers since they replace all zero counts by 0.5 (or they use some other arbitrary constant to impute the zeros), then take a log-ratio transformation of the proportions and apply Euclidean data analysis methods. This is a form of Winsorisation and introduces bias into the estimation procedure.  There are typically huge numbers of zeros in microbiome data and we argue that they should not be automatically treated as outliers, but rather they should be treated as legitimate datapoints that occur with relatively  high probability. 

  The purpose of this article is to develop a novel method of estimation
  for the PPI model with several attractive features: (a) it
  is tractable, (b) it {is insensitive to} zero values in any of the
  components of a data composition, and (c) it is {resistant} to outliers.
Outliers can occur when the majority of the dataset is highly concentrated in a relatively small region of the simplex. An observation not close to the majority would be deemed an outlier.
  The method is based on score matching estimation (SME) after an
  additive log-ratio transformation of the data, plus the inclusion of 
  additional weights in the estimating equations for {resistance to outliers}.
  Although the method is mathematically well-defined for the full PPI
  model, it is helpful in practice to focus on a restricted version of
  the PPI model (the RPPI model) both for identifiability reasons and because the restricted model was shown by Scealy and Wood (2022) to provide a good fit to microbiome data. 

  The article is organized as follows.  The PPI model is presented in Section 2. The important distinction
  between zero components and outliers is explored in Section 3.
  Section 4 gives the score matching algorithm and Section 5 gives
  the modifications needed for {resistance to outliers}.  An application to
  microbiome data analysis is given in Section 6, with additional
  simulation studies in Section 7.  Further technical details are
  given in the Supplementary Material (SM) which also includes a
  document that reproduces the numerical results in this article.

  A package designed for the Comprehensive R Archive Network (CRAN,
  2019) is under development and available at
  \url{github.com/kasselhingee/scorecompdir}. The package
  contains our new additive log-ratio score matching estimator and its robustified version, other score matching estimators, and a
  general capacity for implementing score matching estimators.

  \section{The PPI distribution} \label{sec2:RPPI}
The $(p-1)$-dimensional \textit{simplex} in $\mathbb{R}^p$ is defined by
\begin{equation}
\Delta^{p-1}= \Bigg\{\bu = (u_1,u_2, \ldots,u_p)^{\top} \in \mathbb{R}^p : 
 u_j \geq 0,\quad \sum_{j=1}^pu_j=1  \Bigg\},
\label{simplex}
\end{equation}
where  a \textit{composition} $\bu$ contains $p$ nonnegative \textit{components}
adding to 1.
The boundary of the simplex consists of compositions for which one or
more components equal zero.  The \textit{open simplex}
$\Delta^{p-1}_0$ excludes the boundary so that
$u_j>0, \ j=1, \ldots, p$.

The \textit{polynomially-tilted pairwise interaction (PPI)} model of
Scealy and Wood (2022)
on $\Delta^{p-1}_0$  is defined by the density
\begin{equation}
f(\bu;\bD , \bbeta)=\\
\frac{1}{c_1(\bD,\bbeta)} \left( \prod_{j=1}^p u_j^{\beta_j} \right)
\exp{\left( \bu^\top \bD  \bu \right)}, 
\quad \bu \in \Delta_0^{p-1},
\label{uqdensity1}
\end{equation}
with respect to $d \bu$, where $d \bu$ denotes Lebesgue measure in
$\mathbb{R}^p$ on the hyperplane $\sum u_j =1$.  The density is the
product of a Dirichlet factor and an exp-quadratic (i.e. the
exponential of a quadratic) factor. To ensure integrability, the
Dirichlet parameters must satisfy
$\beta_j>-1, \ j=1, \ldots, p$.  Note that if $-1 < \beta_j<0$, the
Dirichlet factor blows up as $u_j \rightarrow$ 0.  The matrix $\bD$ in
the quadratic form is a symmetric matrix.  Due to the constraint
$\sum u_j =1$ it may be assumed without loss of generality that
$\bone^\top \bD \bone = 0$.

If the last component is written in terms of the earlier components,
$u_p = 1-\sum_1^{p-1} u_j$, then (\ref{uqdensity1}) can be written in
the alternative form
\begin{equation}
f(\bu;\bA_L, \bbeta, \bb_L)=\\
\frac{1}{c_2(\bA_L, \bb_L,\bbeta)} \left( \prod_{j=1}^p u_j^{\beta_j} \right) \exp{\left( \bu_L^\top \bA_L \bu_L + \bu_L^{\top}\bb_L   \right)}, 
\quad \bu \in \Delta_0^{p-1},
\label{uqdensity2}
\end{equation}
with respect to the {same Lebesgue measure $d\bu$}, where $\bu_L=(u_1,u_2,\ldots, u_{p-1})^{\top}$, $\bA_L$ is a
$(p-1) \times (p-1)$-dimensional symmetric matrix and $\bb_L$ is a 
$(p-1)$-dimensional vector.

The full PPI model contains $(p^2+ 3p -2)/2$ parameters.  Although the
parameters are mathematically identifiable, in practice it can be
difficult to estimate all of them accurately.  Hence it is
useful to consider a \textit{restricted PPI (RPPI)} with a smaller
number of free parameters.  The RPPI model contains $q = (p+2)(p-1)/2$
free parameters (the same as for the $(p-1)$-dimensional multivariate
normal distribution) and is defined as follows.  First, order the
components so that the most abundant component $u_p$ is listed last.
Then set
\begin{equation}
  \label{eq:RPPI-constraints}
  \bb_L=0, \quad \beta_p=0.
\end{equation}

\section{Zeros and Outliers}
Two types of extreme behaviour in compositional data are zeros and outliers,
and it is helpful to distinguish between these two concepts.

An outlier is defined to be an observation which has low probability
under the PPI model fitted to the bulk of the data on the 
simplex. Outliers can occur when the majority of the dataset is highly
concentrated in a relatively small region of the simplex, even though
the simplex is a bounded space. In particular, if most of the data are highly
concentrated in the middle of the simplex with small variance, then an
observation close to or on the boundary would be deemed to be an outlier.

On the other hand if the marginal distribution for the $j$th component
has a nonvanishing probability density as $u_j$ tends to 0 (e.g. in the
PPI model with $\beta_j \leq 0$), then a composition $\bu$ with
$u_j=0$ would not be considered to be an outlier.

Although the PPI model has no support on the boundary of the simplex,
we may still want to fit the model to data sets for which some of the
compositions have components which are exact zeros.  There are two
main ways to think about the presence of zeros in the data.  First,
they may be due to measurement error; a measurement of zero
corresponds to a ``true'' composition lying in the interior of the
simplex.  Second the data may arise as counts from a multinomial
distribution where the probability vector is viewed as a latent
composition  coming from the PPI model.  See Sections \ref{consist}, \ref{data1} and Scealy and Wood (2022) for further details on the multinomial latent variable model. Then zero counts can
occur even though the probability vector lies in $\Delta^{p-1}_0$.

The presence of zero components in data poses a major problem for
maximum likelihood estimation for the PPI model.  In particular, the
derivative of the log-likelihood function with respect to $\beta_j$
for a single composition $\bu$,
$$
\partial \log f(\bu;\bD , \bbeta) / \partial \beta_j = \log u_j-{\partial \log{c_1}}/{\partial \beta_j}
$$
is unbounded as $u_j \rightarrow 0$, which leads to singularities in the maximum
likelihood estimates.

Hence we look for alternatives to maximum likelihood estimation.  One promising
general approach is score matching estimation (SME), due to Hyvarinen (2005).
A version of SME was used by Scealy and Wood (2022) that involved downweighting
observations near the boundary of the simplex.  However, their method
was somewhat cumbersome due to the requirement to specify a weight function and their estimator is inefficient when many of the parameters $\beta_j$, $j=1,2,\ldots, p-1$ are close to $-1$. This parameter setting is relevant to microbiome data applications.   

This article uses another version of SME that we call ALR-SME because it involves
a preliminary additive log-ratio transformation of the data.
ALR-SME  is tractable and  is insensitive to zeros, in the sense
that the influence function is bounded as $u_j \rightarrow 0$ for any $j$.

Further, following the method of Windham (1995) it is possible to robustify  ALR-SME {to outliers} by incorporating suitable weights in the estimating equations.  Details
are given in Section \ref{robust_alr}.
{In this article we make a distinction between robustness to zeros and robustness to outliers, and for clarity we often describe these forms of robustness as \textit{insensitive to zeros} and \textit{resistant to outliers}, respectively.}

\section{Additive log-ratio score matching estimation}
In this section we recall the general construction of the score matching estimator
due to Hyvarinen (2005), and then apply it to data from the RPPI distribution
after first making an additive log-ratio transformation.

\subsection{The score matching estimator}
The construction of the score matching estimator starts with the \textit{Hyvarinen divergence}, defined by
\begin{equation}
  \Phi(g,g_0) = \frac12 \int_{\by \in \mathbb{R}^{p-1} }
  \{ \nabla \log g(\by) - \nabla \log g_0(\by)\}^2 g_0(\by) d\by \label{eq:H-div}
\end{equation}
where $g$ and $g_0$ are probability densities on $\mathbb{R}^{p-1}$
subject to mild regularity conditions {(Hyvarinen, 2005)}.  Note that $ \Phi(g,g_0) = 0$
if and only if $g=g_0$.

Let
\begin{equation}
  g(\by) = g(\by; \bpi) \propto \exp\{\bpi^\top \bt(\by)\}
  \label{eq:exp-fam}
\end{equation}
define an exponential family model, where $\bpi$ is a $q$-dimensional
parameter vector and $\bt(\by)$ is a $q$-dimensional vector of
sufficient statistics.  Then for a given density $g_0$, the
``best-fitting'' model $g(\by; \bpi)$ to $g_0$ can be defined by
minimizing (\ref{eq:H-div}) over $\bpi$.  Since $\nabla \log g(\by) $
is linear in $\bpi$, $\Phi$ is quadratic in $\bpi$.  Differentiating
$\Phi$ with respect to $\bpi$, and setting the derivative to $\bzero$ yields
the estimating equations
$$
\bW \bpi -   \bd = \bzero,
$$
where $\bW$ and $\bd$ have elements
\begin{align}
  w_{k_1,k_2} &= \int \sum_{j=1}^{p-1}
                (\partial t_{k_1}(\by) / \partial y_j) 
                (\partial t_{k_2}(\by) / \partial y_j)
                g_0(\by) d\by, \quad k_1,k_2=1, \ldots, q,
  \label{eq:W}\\
  d_k &= \Delta t_k(\by) = \int \sum_{j=1}^{p-1}
        (\partial^2 t_k(\by) / \partial y_j^2)
        g_0(\by) d\by,
  \quad k=1, \ldots, q. \label{eq:d}
\end{align}
The Laplacian in (\ref{eq:d}) arises after integration by parts in (\ref{eq:H-div}).
Hence the the best-fitting value of $\bpi$ is 
$$
\bpi = \bW^{-1}  \bd.
$$

Given data $\by_i, \ i=1, \ldots, n$, with elements
$y_{ij}, \ j=1, \ldots, p-1$,
the integrals can be replaced by empirical averages to yield the
estimating equations
\begin{equation}
  \label{eq:sme-esteqn}
\hat{\bW} \bpi -   \hat{\bd} = \bzero,
\end{equation}
where $\hat{\bW}$ and $\hat{\bd}$ have elements
\begin{align}
  \hat{w}_{k_1,k_2} &= \sum_{i=1}^n \sum_{j=1}^{p-1}
              (\partial t_{k_1}(\by_i) / \partial y_j) 
              (\partial t_{k_2}(\by_i) / \partial y_j),
              \quad k_1,k_2=1, \ldots, q,  \label{eq:W-hat}\\
  \hat{d}_k &= \Delta t_k(\by) = \sum_{i=1}^n \sum_{j=1}^{p-1}
        (\partial^2 t_k(\by_i) / \partial y_j^2),
        \quad k=1, \ldots, q. \label{eq:d-hat}
\end{align}
Solving the estimating equations (\ref{eq:sme-esteqn}) yields the score matching estimator (SME)
\begin{equation}
  \label{eq:sme}
\hat \bpi = \hat{\bW}^{-1} \hat{\bd}.
\end{equation}

\subsection{Additive log-ratio transformed compositions}

To make use of this result for  distributions on the simplex, it is helpful to make
an initial additive log-ratio (ALR) transformation from $\bu \in \Delta^{p-1}_0$
to $\by = (y_1,y_2,\ldots, y_{p-1})^{\top} \in \mathbb{R}^{p-1}$ where
$$
y_j = \log(u_j/u_p), \quad j=1, \ldots, p-1.
$$
This transformation was popularized by Aitchison (1986).  The logistic-normal distribution for $\bu$, or equivalently the normal distribution for $\by$ has often been suggested as a model for compositional data (e.g., Aitchison, 1986). However, it should
be noted that the RPPI distribution has very different properties. In particular, we do not advocate  the use of logistic-normal models in situations where zero or very-near-zero compositional components occur frequently. See Scealy and Welsh (2014) for relevant discussion.

The transformed
RPPI distribution has density proportional to
\begin{equation}
  \exp{\left(\frac{\exp{(\by)}^{\top} \bA_L \exp{(\by)}}
      {(1+\sum_{k=1}^{p-1}\exp{(y_k)})^{2}} \right)} \times
  \left( \frac{1}{1+\sum_{k=1}^{p-1}\exp{(y_k)}} \right)
  \prod_{j=1}^{p-1} \left(\frac{\exp(y_j)}{1+\sum_{k=1}^{p-1}\exp{(y_k)}}
  \right)^{\beta_j+1}
\label{lograt}
\end{equation}
with respect to Lebesgue measure $d\by = dy_1 \cdots dy_{p-1}$ on $R^{p-1}$, 
where
$$
\exp{(\by)}=(\exp(y_1),\exp(y_2),\ldots, \exp(y_{p-1}))^{\top}
$$
and we have used the constraints (\ref{eq:RPPI-constraints}).
The density (\ref{lograt})  forms a canonical exponential family with
parameter vector 
\begin{equation*}
\bpi=(a_{11},a_{22}, \ldots, a_{(p-1)(p-1)}, a_{12},a_{13},\ldots, a_{(p-2)(p-1)},
1+\beta_1, 1+\beta_2\ldots, 1+\beta_{p-1})^{\top}
\end{equation*} 
with $q={p(p-1)}/{2} + (p-1)$ parameters, {where $a_{ij}$ refers to the $i$, $j$th element of $\bA_L$}.  The elements of
$\hat{\bW}$ and $\hat{\bd}$ in (\ref{eq:sme-esteqn})
can be expressed in terms of linear
combinations of powers and products of the $u_{ij}$ which are the elements of the data vectors $\bu_i$, $i=1,2,\ldots, n$; see the equations
(\ref{Wmom}) and (\ref{dmom}) in Appendix \ref{details} (SM).  We refer
to the resulting score matching estimator  as the ALR-SME.

\subsection{Consistency \label{consist}}
Next we state a consistency result for the ALR-SME when applied to the
multinomial latent variable model.  Let $\bx_i, \ i=1, \ldots, n$, be
independent multinomial count vectors from different multinomial
distributions, where the probability vectors $\bu_i$ are taken
independently from the RPPI model.  Let
$m_i = x_{i1} + \cdots + x_{ip}$ denote the total count from the $i$th
multinomial vector.  That is, we assume the conditional probability mass function of $\bx_i=(x_{i1},x_{i2},\ldots, x_{ip})^{\top} $ given $\bu_i$ is  
$f(\bx_i \vert \bu_i)=m_i! \prod_{j=1}^p \{ u_{ij}^{x_{ij}}/x_{ij}!\}$, where the $\bu_i=(u_{i1},u_{i2},\ldots, u_{ip})^{\top}$ are unobserved latent variables. This model is relevant for analysing microbiome data; see Section \ref{data1}.
Consider estimating  the parameters $\bA_L$ and
$\beta_1, \beta_2, \ldots, \beta_{p-1}$ using the ALR-SME where the
known proportions $\hat{\bu}_i=\bx_i/m_i$ are used as substitutes for
the unknown true compositions $\bu_i$ for $i=1,2,\ldots, n$. Note that
we do not need the extra restrictive conditions in part (III) Theorem
3 of Scealy and Wood (2022) for estimating $\bbeta$. The proof of
Theorem 1 below is very similar to the proof of Theorem 3 in Scealy
and Wood (2022) and is omitted.

\noindent \textbf{Theorem 1.}  \textit{Let $\hat{\bpi}$ denote the
  ALR-SME of $\bpi$ (\ref{eq:sme})  based on the (unobserved)
  compositional vectors $\bu_1, \ldots , \bu_n$ and let $\hat{\bpi}^{\dagger}$
  denote the ALR-SME of $\bpi$ based
  on the observed vectors of proportions $\hat{\bu}_1, \ldots , \hat{\bu}_n$.
  Assume that for some constants $C_1>0$ and $\alpha>1$,
  $\inf_{i=1, \ldots , n} m_i \geq C_1 n^\alpha$.\\
  Then as $n \rightarrow \infty$
\begin{equation}
  \vert \vert n^{1/2}\left (\hat{\bpi}^{\dagger} - \hat{\bpi} \right )\vert \vert
  =o_p(1).
\label{convergenceT}
\end{equation} 
}

Theorem 1 above shows that $\hat{\bpi}^{\dagger}$ is asymptotically
equivalent to $\hat{\bpi}$ to leading order. Note that Theorem 1 does
not assume that the population latent variable distribution is a RPPI
distribution, but if the RPPI model is correct then
$\hat{\bpi}^{\dagger}$ and $\hat{\bpi}$ are both consistent estimators
of $\bpi$ under the conditions of Theorem 1. Asymptotic normality of
$\hat{\bpi}$ also follows directly from similar arguments to
Scealy and Wood (2022). Theorem 1 applies even when the observed data
has a large proportion of zeros. This has important implications for
analysing microbiome count data with many zeros. Scealy and Wood
(2022) were unable to use score matching based on the square root
transformation to estimate $\bbeta$ when analysing real microbiome
data because the extra conditions needed for consistency did not look
credible for the real data (there was an extra assumption needed on
the marginal distributions of the components of the $\bu_i$). Here,
using the ALR-SME we are now able to estimate $\bbeta$
directly using score matching. See Sections \ref{data1} and
\ref{simulation_sec} for further details.

It is also insightful to compare the ALR-SME to standard maximum likelihood estimation. The maximum
likelihood estimator for $\bA_L$ and $\bbeta$ based on an iid sample
from model (\ref{uqdensity2}) solves the estimating equation
\begin{equation}
  \sum_{i=1}^n \left( \bt (\bu_i) -\frac{\partial}{\partial \bpi}
    \log{\{c_2(\bA_L, \boldsymbol{0},\bbeta)\}}\right)=\boldsymbol{0},
\label{ml1}
\end{equation}
where $\bt(\bu)$ is defined at (\ref{tdef}) in Appendix
\ref{details} (SM). Denote the maximum likelihood estimator of $\bpi$ by
$\hat{\bpi}_{ML}$. The estimator $\hat{\bpi}_{ML}$ is difficult to
calculate due to the intractable normalising constant $c_2$. Theorem 1
also does not hold for $\hat{\bpi}_{ML}$ due to the presence of the
$\log{(u_j)}$ terms in $\bt (\bu)$ which are unbounded at zero. That
is, we cannot simply replace $\bu_i$ by $\hat{\bu}_i$ within
(\ref{ml1}) to obtain a consistent estimator for the multinomial
latent variable model. This is a major advantage of the ALR-SME
because it leads to computationally simple and consistent estimators,
whereas $\hat{\bpi}_{ML}$ with the latent variables $\bu_i, \ i=1,2,\ldots, n$ each replaced with $\hat{\bu}_i$  is inconsistent and
computationally not tractable.

\subsection{Comments on SME}
Score matching estimation has been defined here for probability
densities whose support is all of $\mathbb{R}^d$.  This construction can
be extended in various ways, and we mention two possibilities here that are
relevant for compositional data.

First, the unbounded region $\mathbb{R}^{p-1}$ in (\ref{eq:H-div}) can be
replaced by a bounded region such as $\Delta^{p-1}_0$.  However, there is a price to pay.  The
integration by parts which underlies the Laplacian term in (\ref{eq:d}) now
includes boundary terms.  Scealy and Wood (2022) introduced a
weighting function which vanishes on the boundary of the
simplex.  The effect of this weighting function is to eliminate the
boundary terms.  However, the weighting function also lessens the
contribution of data near the boundary to the estimating equations.

Second, the Hyvarinen divergence in (\ref{eq:H-div}) implicitly uses a Riemannian
metric in $\mathbb{R}^{p-1}$, namely Euclidean distance.  Other choices
of Riemannian metric lead to different estimators.  Some comments on these
choices in the context of compositional data are discussed in Appendix \ref{metric_score} (SM).

\section{An ALR-SME {that is Resistant to Outliers} \label{robust_alr}}

Although the simplex is a bounded space, outliers/influential points
can still occur when the majority of the data is highly concentrated
in certain regions of the simplex. In the case of microbiome data (see
Section \ref{data1}), there are a small number of abundant components
which have low concentration (e.g. {\it Actinobacteria} and {\it
  Proteobacteria}) and these components should be fairly resistant to
outliers. However, the components {\it Spirochaetes}, {\it
  Verrucomicrobia}, {\it Cyanobacteria/Chloroplast} and {\it TM7} are
highly concentrated at or near zero and any large values away from
zero can be influential. For the highly concentrated microbiome
components distributed close to zero, these marginally look to be
approximately gamma or generalised gamma distributed; see Figures
\ref{figa} and \ref{figb} in Section \ref{data1}.  Hence there is a
need for the Dirichlet component of the density in the RPPI model
\eqref{uqdensity2}.

We now develop score matching estimators for the RPPI model
\eqref{uqdensity2} {that are resistant to outliers}. Assume
that the first $k^*$ components of $\bu$ are highly concentrated near
zero where it is expected that possibly
$\beta_1 < 0, \beta_2 < 0, \ldots, \beta_{k^{*}} < 0$.  The remaining
components $u_{k^*+1}, u_{k^*+2},\ldots, u_p$ are assumed to have
relatively low concentration. By low concentration we mean moderate to high variance and by high concentration we mean small variance. The robustification which follows is only relevant for highly concentrated components near zero which is why we are distinguishing between the different cases. See Scealy and Wood (2021) for further discussion on standardised bias robustness under high concentration which is relevant to all compact sample spaces including the simplex. 
When $k^* < p-1$
partition
\begin{equation*}
\bA_L=
\begin{pmatrix}
\bA_{KK} & \bA_{KR} \\
\bA_{RK} & \bA_{RR}
\end{pmatrix},
\end{equation*}
where $\bA_{KK}$ is a $k^*\times k^*$ matrix, $\bA_{KR}$ is a
$k^* \times (p-1-k^*)$ matrix, $\bA_{RK}$ is a $(p-1-k^*) \times k^*$
matrix and $\bA_{RR}$ is a $(p-1-k^*) \times (p-1-k^*) $ matrix. When
$k^*=p-1$ then $\bA_L=\bA_{KK}$.  The (unweighted) estimating
equations for the ALR-SME are given by (\ref{eq:sme-esteqn}) and can be written
slightly more concisely as 
\begin{equation}
\boldsymbol{0}=
\frac1n\sum_{i=1}^n \left( \bW_1(\bu_i) \bpi - \bd_1(\bu_i) \right),
\label{unweighted}
\end{equation}
where the elements of $\bW_1(\bu_i)$ and $\bd_1(\bu_i)$ are functions of $\bu_i$ and are defined at equations (\ref{Wmom}) and (\ref{dmom}) in Appendix \ref{details} (SM) for the RPPI model and are given in a more general form though equations (\ref{eq:sme-esteqn})-(\ref{eq:d-hat}).

Windham's (1995) approach to creating robustified estimators is to use
weights which are proportional to a power of the probability density
function.  The intuition behind this approach is that outliers under a
given distribution will typically have small likelihood and hence a
small weight, whereas observations in the central region of the
distribution will tend to have larger weights.  The Windham (1995)
method is an example of a density-based minimum divergence estimator,
but with the advantage that the normalising constant in the density
does not need to be evaluated in order to apply it.  See Windham
(1995), Basu et al. (1998), Jones et al. (2001), Choi et al. (2000),
Ribeiro and Ferrari (2020), Kato and Eguchi (2016) and Saraceno et
al. (2020) for further discussion and insights.  In the setting of the RPPI model for compositional data,  there
is a choice to be made between the probability densities to use in the
weights, that is to use (\ref{uqdensity2}) or (\ref{lograt}), or in
other words should we choose the measure $d\bu$ or $d \by$. We prefer
$d \bu$ because $d \by$ places zero probability density at the simplex
boundary and thus always treats zeros as outliers which is not a good property with data concentrated near the simplex boundary.

For the RPPI distribution, taking a power of the density
(\ref{uqdensity2}) is a bad idea because for those $\beta_j$ which are
negative the weights will be infinite as $u_j$ tends to $0$.  To
circumvent this issue we only use the exp-quadratic factor in
(\ref{uqdensity2}) to define the weights.  This choice of weighting
function is a compromise between wanting the weight of an observation
to be smaller if the probability density is smaller and needing to
avoid infinite weights on the boundary of the simplex.  
{In fact, 
typically $\bu_{K}^\top \bA_{KK}\bu_{K}$, where $\bu_{K}=(u_{1},u_{2},\ldots, u_{k^*})^{\top}$, is highly negative whenever $\bu$ has a large value in any of the components that are highly concentrated near zero in distribution.
It is thus sufficient to use just the $\exp(\bu_{K}^\top \bA_{KK}\bu_{K})$ factor of (\ref{uqdensity2}) in the weights (the influence function in Theorem 2 below confirms this behaviour).
Including all elements of $\bA_L$ in the weights leads to a large loss in efficiency,
so the weights in our robustified ALR-SME estimator are  
$\exp(c\bu_{i,K}^\top \bA_{KK}\bu_{i,K})$, $i=1, \ldots , n$, where
$\bu_{i,K}=(u_{i1},u_{i2},\ldots, u_{ik^*})^{\top}$.
}
The weighted form of estimating equations
(\ref{unweighted}) is then
\begin{equation}
\sum_{i=1}^n \exp{\left( c\bu_{i,K}^\top \bA_{KK} \bu_{i,K}   \right)} \left( \bW_1(\bu_i) \bH \bpi - \bd_1(\bu_i) \right)=\boldsymbol{0},
\label{windham2}
\end{equation}
where $\bH$ is a $q \times q$ diagonal matrix with diagonal
elements either equal to $c+1$ or 1 (the elements corresponding to the
parameters $\bA_{KK}$ are $c+1$ and the rest are 1).
The estimating
equation (\ref{windham2}) has a very simple form here (i.e. given the weights, the estimating equations are linear in $\bpi$), whereas the
version based on the maximum likelihood estimator does not have such a
nice linear form leading to a much more complicated influence function calculation and its interpretation (e.g. Jones et al., 2001).

An algorithm similar to that in Windham (1995) can be used to solve
(\ref{windham2}) and involves iteratively solving weighted versions of
the score matching estimators. In summary this algorithm is
\begin{itemize}
\item [1.] Set $r=1$ and initialise the parameters:
  $\hat{\bbeta}^{(0)}$ and $\hat{\bA}_L^{(0)}$ (i.e. choose starting
  values such as the unweighted ALR-SME). Then repeat steps 2-5 until convergence.
\item [2.] Calculate the weights
  $ \tilde{w}_i= \exp{\left( c\bu_{i,K}^\top \hat{\bA}_{KK}^{(r-1)}
      \bu_{i,K} \right)}$ for $i=1,2,\ldots, n$ and normalise the
  weights so that the weights sum to 1 across the sample. Also
  calculate the additional tuning constants
  $\bd_{\beta}=-c\hat{\bbeta}^{(r-1)}$,
  $\bd_{A_{1}}=-c\hat{\bA}_{RR}^{(r-1)}$,
  $\bd_{A_{2}}=-c\hat{\bA}_{RK}^{(r-1)}$ and
  $\bd_{A_{3}}=-c\hat{\bA}_{KR}^{(r-1)}$.
\item [3.] Calculate weighted score matching estimates. That is,
replace all sample averages with weighted averages using the
normalised weights $\tilde{w}_i$ calculated in step 2. Denote the
resulting estimates as $\tilde{\bbeta}^{(r)}$ and
$\tilde{\bA}_L^{(r)}$.
\item [4.] The estimates in step 3 are biased and we need to do the
following bias correction:
\begin{equation*} \hat{\bbeta}^{(r)}=
\frac{\tilde{\bbeta}^{(r)}-\bd_{\beta}}{c+1}, \quad \text{and} \quad
\hat{\bA}_{KK}^{(r)}=\frac{\tilde{\bA}_{KK}^{(r)}}{c+1}
\end{equation*} and
\begin{equation*}
\hat{\bA}_{RR}^{(r)}=\frac{\tilde{\bA}_{RR}^{(r)}-\bd_{A_1}}{c+1},
\quad
\hat{\bA}_{RK}^{(r)}=\frac{\tilde{\bA}_{RK}^{(r)}-\bd_{A_2}}{c+1},
\text{and} \quad
\hat{\bA}_{KR}^{(r)}=\frac{\tilde{\bA}_{KR}^{(r)}-\bd_{A_3}}{c+1}.
\end{equation*} This correction is simple because the model is an
exponential family; see Windham (1995) for further details.
\item [5.] $r \rightarrow r+1$
\end{itemize}

Step 4 in this new robust score matching algorithm above is similar to
applying the inverse of $\tau_c$ in Windham (1995). The tuning
constants $\bd_{\beta}$, $\bd_{A_1}$, $\bd_{A_2}$ and $\bd_{A_3}$ are
required due to our use of a factor of the density in the weights.

This modified version of Windham's (1995) method is particularly
useful when any $\beta_j$'s are negative in order to avoid infinite
weights at zero. When the data is concentrated in the
simplex interior (i.e. we expect $\beta_j > 0$, $j=1,2,\ldots, p$) then
the model density is bounded and we can apply the Windham
(1995) method without modification{, although efficiency gains may be possible from using well-chosen factors of the model density}.

In order to complete the description of the robustified ALR-SME,  we need to
choose the robustness tuning constant $c$. In related settings Kato
and Eguchi (2016) use cross validation and Saraceno et al. (2020)
calculate theoretical optimal values for a Gaussian linear mixed
model. Basak et al. (2021) report that choosing the optimal tuning
constant is challenging in general when choosing density power
divergence tuning parameters. We agree with the view of Muller and
Welsh (2005, page 1298) that choice of model selection criteria or
estimator selector criterion should be independent of the estimation
method, otherwise we may excessively favour particular
estimators. This is an issue with the Kato and Eguchi (2016) method
which is based on an arbitrary choice of divergence which could favor
the optimal estimator under that divergence. Instead we use a
simulation based method to choose $c$; see Section \ref{data1}. We
also need to decide on the value of $k^*$; see Section \ref{data1} for
a guide.

We next examine the theoretical properties of our new robust{ified}
estimator.  Let $\mathcal{F}$ denote the set of probability
distributions on the unit simplex $\Delta^{p-1} \subset \mathbb{R}^p$,
where $p \geq 3$. Let $F_0$ denote the population probability measure
for a single observation from $\Delta^{p-1} $ and write $\delta_{\bz}$
for the degenerate distribution on $\Delta^{p-1} $ which places unit
probability on $\bz \in \Delta^{p-1} $. Consider the ALR-SME  functional
$\btheta: \mathcal{F} \rightarrow \Theta \subseteq \mathbb{R}^{q}$. It
is assumed that $\btheta$ is well defined for all $\bz$ at
$(1-\lambda)F_0+ \lambda \delta_{\bz}$ provided $\lambda \in (0,1)$ is
sufficiently small. Then the influence function for
$\btheta$ and $F_0\in \mathcal{F}$ at $\bz$ is defined by
\begin{equation*} \text{\bf IF}_{\btheta;F_0} (\bz)=\lim_{\lambda
\rightarrow 0} \frac{1}{\lambda} \left( \btheta \left\{
(1-\lambda)F_0+ \lambda \delta_{\bz} \right\} -\btheta \left( F_0
\right) \right).
\end{equation*}

\noindent \textbf{Theorem 2.}  \textit{ Suppose that the population
distribution $F_0$ on $\Delta^{p-1} $ is absolutely continuous with
respect to Lebesgue measure on $\Delta^{p-1}$. Also assume that
$k^*=p-1$ for exposition simplicity which implies that all of the first $p-1$ components are concentrated at/near zero.  (The proof for the case of $k^*
< p-1$ is similar and is not presented here.) Then
\begin{equation*} \text{\bf IF}_{\bpi;F_0}
(\bz)=-\left(\bG(\bpi_0)\right)^{-1} \exp{\left( c
\bt^{(a)}(\bz)^{\top} \bpi_0\right)} \left \{ \bW_1(\bz) \bH \bpi_0
-\bd_1 (\bz) \right \},
\end{equation*} where $\bpi_0$ is the solution to the population
estimating equation {corresponding to (\ref{windham2})} (see equation (\ref{pop}) in Appendix
\ref{influence} (SM)) and the functions $\bG(\bpi_0)$ and $\bt^{(a)}(\bz)$
are defined in Appendix \ref{influence} (SM).  }

The functions $\bt^{(a)}(\bz)$, $\bW_1(\bz)$ and $\bd_1(\bz)$ contain
linear combinations of low order polynomial products, for example
terms like $z_{1}^{r_1}z_{2}^{r_2}z_{3}^{r_3}$, where $r_1 \geq 0$,
$r_2 \geq 0$, $r_3 \geq 0$ and $r_1+r_2+r_3$ is small. Therefore the
above influence function is always bounded for all $\bz \in
\Delta^{p-1}$ including for any points on the simplex boundary{, even when c = 0}. 
{The $\bt^{(a)}(\bz)^{\top} \bpi_0$ in Theorem 2 is equal to $\bz_K^{\top} \bA_{KK} \bz_K$, where $\bz_K = (z_1, z_2, ..., z_{k^*})^\top$. 
For many PPI
models, {$\bz_K^{\top} \bA_{KK} \bz_K=$} $\bt^{(a)}(\bz)^{\top} \bpi_0 < 0$, which means that for the
components of $\bu$ that are highly concentrated near zero in
distribution, any large value away from zero in these components will
be down-weighted and have less influence on the estimator. This leads
to large efficiency gains in both the contaminated and uncontaminated
cases. See Section \ref{simulation_sec} for further details. 

It is useful to compare Theorem 2 with the influence function for
$\hat{\bpi}_{ML}$. The maximum likelihood estimator is a standard
M-estimator with influence function of the form
\begin{equation}
-\bB^{-1} \left( \bt (\bz) -\frac{\partial}{\partial \bpi_0} \log{c_2}\right),
\label{influenceML}
\end{equation}
where $\bB$ is a matrix function of the model parameters (e.g. Maronna
et al., 2006, page 71) and $\bt(\bz)$ is defined at (\ref{tdef}) in
Appendix \ref{details} (SM). The vector $t(\bz)$ contains the functions
$\log(z_1), \ \log(z_2), \ \ldots, \ \log(z_{p-1})$ and the influence
function (\ref{influenceML}) is unbounded if any $z_j$ approaches $0$,
$j=1,2,\ldots, p-1$.  Therefore maximum likelihood estimation for the PPI model {is highly sensitive to zeros}.
Maximum likelihood estimation is also {highly sensitive to zeros} for
the gamma, Beta, Dirichlet and logistic normal distributions for
similar reasons.

\section{Microbiome data analysis \label{data1}}

Microbiome data is challenging to analyse due to the presence of high right skewness, outliers and zeros in the marginal distributions of the bacterial species (e.g. Li, 2015; He et al, 2021).  Typically microbiome count data is either modelled using a multinomial model with latent variables (e.g. Li, 2015; Martin et al., 2018; Zhang and Lin, 2019) or the sample counts are normalised and treated as approximately continuous data since the total counts are large (e.g. Cao et al., 2019; He et al., 2021). Here we analyse real microbiome count data by fitting a RPPI multinomial latent variable model using the normalised {microbiome} counts as estimates of the latent variables; see Section \ref{consist}.   

In this section we analyse a subset of the longitudinal microbiome dataset obtained from a study carried out in a helminth-endemic area in Indonesia (Martin et al., 2018). In summary, stool samples were collected from 150 subjects in the years 2008 (pre-treatment) and in 2010 (post-treatment). The 16s rRNA gene from the stool samples was processed and resulted in counts of 18 bacterial phyla. Whether or not an individual was infected by helminth was also determined at both time points.  We restricted the analysis to the year 2008 for individuals infected by helminths which resulted in a sample size of $n=94$, and we treated these individuals as being independent. 

Martin et al. (2018) analysed the five most prevalent phyla and pooled the remaining categories. Scealy and Wood (2022) analysed a different set of four phyla including two with a high number of zeros and pooled the remaining categories. Here for demonstrative purposes we will first analyse the same data components as in Scealy and Wood (2022) with the $p=5$ components representing {\it TM7}, {\it Cyanobacteria/Chloroplast}, {\it Actinobacteria}, {\it Proteobacteria} and {\it pooled}. The percentage of zeros in each category are $38\%$, $41\%$, $0\%$, $0\%$  and  $0\%$ respectively. Call this {\it Dataset1}. Then for demonstrative purposes we will also analyse a second dataset with $p=5$ denoted as {\it Dataset2} which contains the components {\it Spirochaetes}, {\it Verrucomicrobia}, {\it Cyanobacteria/Chloroplast}, {\it TM7} and {\it pooled}.  The percentage of zeros in each category for {\it Dataset2} are  $77\%$, $75\%$, $41\%$, $38\%$ and $0\%$ respectively. Let $x_{ij}$, $i=1,2,\ldots, 94$ and $j=1,2,3,4,5$ represent the sample counts for a given dataset with total count $m_i=2000$. 
The estimated sample proportions were calculated as follows:
$\hat{u}_{ij}=x_{ij}/m_i$, where $i=1,2,\ldots, 94$ and $j=1,2,3,4,5$.

\begin{figure*}
\centering
\includegraphics[trim = 0cm 0cm 0cm 0cm, clip, height=12cm,width=12cm]{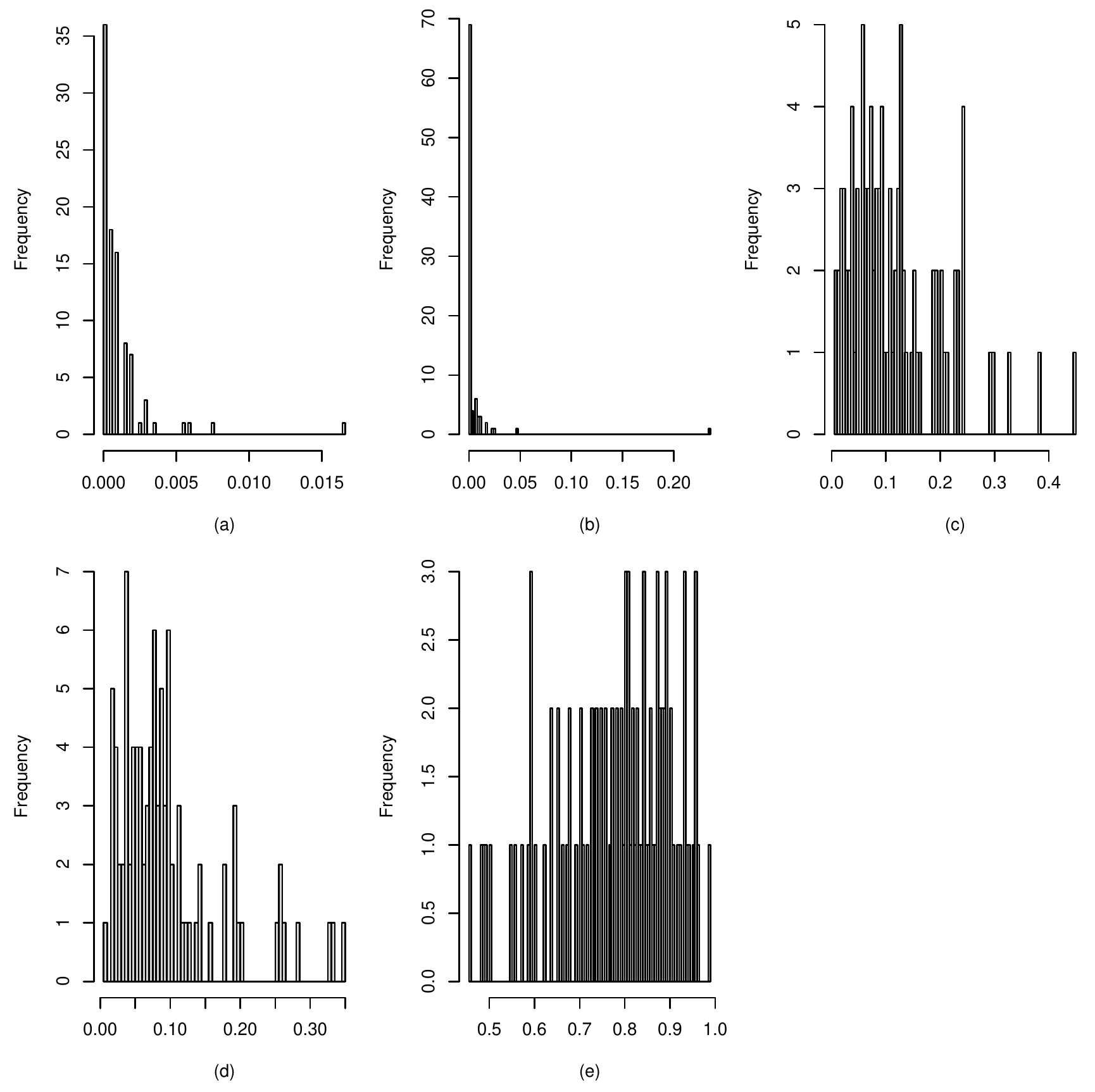}
\caption{{\it Dataset1} histograms of sample proportions $\hat{u}_{ij}$, $j=1,2,3,4, 5.$}
\label{figa}
\end{figure*}

\begin{figure*}
\centering
\includegraphics[trim = 0cm 0cm 0cm 0cm, clip, height=12cm,width=12cm]{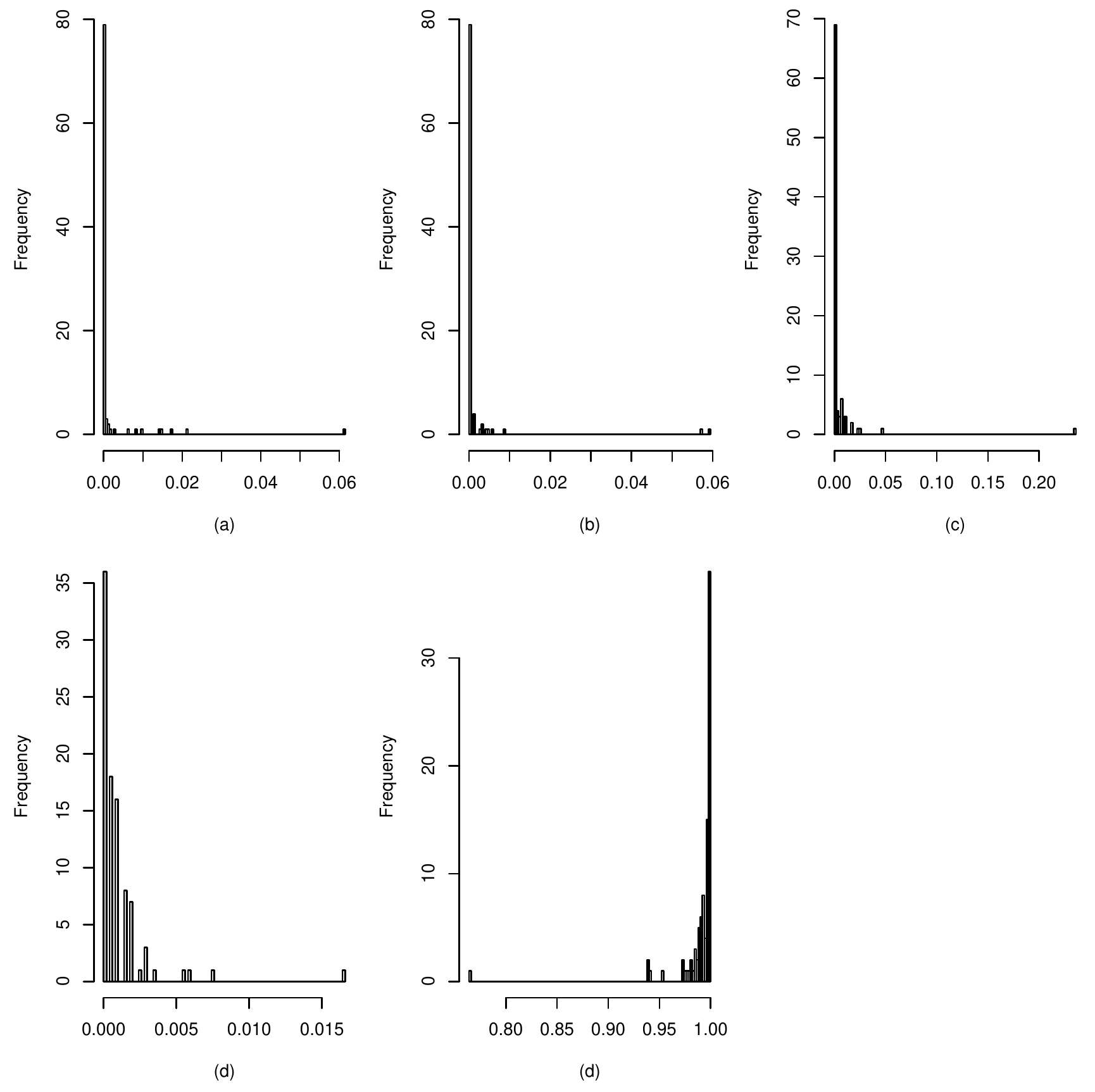}
\caption{{\it Dataset2} histograms of sample proportions $\hat{u}_{ij}$, $j=1,2,3,4, 5.$}
\label{figb}
\end{figure*}

Figure \ref{figa} is similar to Scealy and Wood (2022, Figure 3), the only difference being that we have now included the two large proportions in $\hat{u}_{i1}$ and $\hat{u}_{i2}$ which were deleted by Scealy and Wood (2022) prior to their analysis because they identified them as outliers. The estimates of $\beta_1$ and $\beta_2$ in Scealy and Wood (2022) were negative and close to $-1$ and the components  $\hat{u}_{i1}$ and $\hat{u}_{i2}$ are highly concentrated mostly near zero. The components $\hat{u}_{i3}$ and $\hat{u}_{i4}$ for this dataset have low concentration. Therefore it makes sense here to choose $k^*=2$.  

Figure \ref{figb} contains histograms of the sample proportions in {\it Dataset2}. The first four components are highly concentrated near zero and we would expect that $\beta_1$, $\beta_2$, $\beta_3$ and $\beta_4$ are negative. Therefore it makes sense to choose $k^*=4$ for this dataset.

\subsection{Choice of tuning constant $c$}
For each dataset we let $c$ range over a grid from $0$ up to $1.5$ and we fitted the model for each value of $c$. We simulated a single large sample of size $R=10,000$ under the fitted model (\ref{uqdensity2}) for each value of $c$ and rounded the simulated data as follows: $\hat{u}_{ij}^r=\text{round}(\hat{u}_{ij}m_i)/m_i$, where $\hat{u}_{ij}$ denotes the simulated proportion under the fitted RPPI model for $i=1,2,\ldots, R$. This mimics the discreteness in the data; see Scealy and Wood (2022) Section 7.  Then we compared the simulated proportions with the true sample proportions. Similar to the view of Muller and Welsh (2005, page 1298) we are interested in fitting the core of the data and we are not specifically interested in fitting in the upper tails which is where outliers can occur in this setting. This means we need to choose a criterion that is not sensitive to the upper tail. When comparing the simulated proportions with the true sample proportions we deleted all observations above the $95\%$ quantile cutoff in the marginal distribution proportions.  

For each dataset, $c$ was chosen to give a compromise between fitting all components to give a small value of the Kolmogorov-Smirnov test statistic and keeping variation in the weights as small as possible to preserve efficiency. See Table \ref{tab1} for the chosen values of $c$ for each dataset. Note that the $p$-value for {\it Proteobacteria} is quite small. This is not surprising as this was also the worst fitting component in Table 5 in Scealy and Wood (2022) in their analysis.

\begin{table*}[h]
\centering
{ \footnotesize
\caption{Kolmogorov-Smirnov test results (upper $5\%$ quantile removed)}
\label{tab1}
\begin{tabular}{|c|c|c|c|}
\hline
{\it Dataset1} & $c=0.7$  & {\it Dataset2}  &  $c=1.25$ \\
\hline
{\it TM7} & 0.14  ($p$-value=0.07) &  {\it Spirochaetes}  & 0.088 ($p$-value=0.50)  \\
{\it Cyanobacteria/Chloroplast} & 0.13  ($p$-value=0.11)   &  {\it Verrucomicrobia} & 0.053 ($p$-value=0.96)  \\
{\it Actinobacteria} &0.11 ($p$-value=0.24)  & {\it Cyanobacteria/Chloroplast} & 0.058 ($p$-value=0.93)  \\
 {\it Proteobacteria} & 0.16 ($p$-value= 0.021)    &  {\it TM7}    & 0.058 ($p$-value=0.93)  \\
\hline
\end{tabular}
}
\end{table*}

Table \ref{tab2} contains the parameter estimates for {\it Dataset1}. The standard errors (SE) were calculated using a parametric bootstrap by simulating under the fitted multinomial latent variable model. Use of robust non-parametric bootstrap methods such as those in Muller and Welsh (2005) and Salibian-Barrera et al. (2008) is challenging here due to the large numbers of zeros in the data and we prefer the parametric bootstrap.  The parametric bootstrap SE estimates are expected to be a little larger than the ones in Scealy and Wood (2022) since they used asymptotic standard errors which tended to be a slight underestimation as shown in their simulation study. The new robustified ALR-SME of $\bbeta$ are reasonably close to the simulation/grid search estimates in Scealy and Wood (2022). Interestingly, $\beta_3$ is not significantly different from zero; Scealy and Wood (2022) set this parameter to zero based on visual inspection of plots. As expected the estimates of $\beta_1$ and $\beta_2$ are negative and are highly significant. We no longer need to treat $\bbeta$ as a tuning constant and we can now estimate its standard errors which is an advantage of the new robustified ALR-SME method.

\begin{table*}[h]
\centering
{ \footnotesize
\caption{Parameter estimates and standard errors for {\it Dataset1}}
\label{tab2}
\begin{tabular}{|c|c|c||c|c|c|}
\hline
 parameter & estimate &  estimate/SE & parameter  & estimate &  estimate/SE \\
\hline
$a_{11}$ & -60529.900  & -0.382  & $a_{23}$ & 73.34150      & 0.254 \\
$a_{12}$ &  12432.5000  & 0.255  & $a_{24}$ & 84.06690     & 0.356    \\
$a_{13}$ &  430.21700   & 0.546   & $a_{33}$ & -22.03470   &  -2.62 \\
$a_{14}$ & -411.20500  &   -0.466  & $a_{34}$ &  -9.08582   &  -0.861   \\
$a_{22}$ &  -4934.7000   & -0.303  & $a_{44}$ &  -22.16840  & -2.33 \\
\hline
$\beta_1$ & -0.770598 & -10.8 & $\beta_{3}$ & -0.079014 &  -0.408  \\
$\beta_2$ & -0.870535 & -15.0  & $\beta_{4}$ & -0.149064  & -0.878  \\
\hline
\end{tabular}
}
\end{table*}

Table \ref{tab3} contains the parameter estimates for {\it Dataset2}. Note that we cannot apply the score matching estimators of Scealy and Wood (2022) to this dataset as every datapoint has a component equal to zero, which means the manifold boundary weight functions in Scealy and Wood (2022) evaluate to zero.
However, our new robustified ALR-SME method can handle this dataset with massive numbers of zeros. Again, we used the parametric bootstrap to calculate the standard error estimates. As expected all of $\beta_1$, $\beta_2$, $\beta_3$ and $\beta_4$ are negative and highly significantly different from zero. The $\bA_L$ parameter estimates are all insignificant. So for this dataset perhaps a Dirichlet model might be appropriate. Note that this is not surprising due to the massive numbers of zeros and relatively small sample size; there is not much information available to estimate $\bA_L$.

\begin{table*}[h]
\centering
{ \footnotesize
\caption{Parameter estimates and standard errors for {\it Dataset2}}
\label{tab3}
\begin{tabular}{|c|c|c||c|c|c|}
\hline
 parameter & estimate &  estimate/SE & parameter  & estimate &  estimate/SE \\
\hline
$a_{11}$ & -141.924  & -0.033 & $a_{23}$ & -38106.8     & -0.668 \\
$a_{12}$ &  -16586 & -0.170  & $a_{24}$ & 11709.2       & 0.062    \\
$a_{13}$ & -5877.63   & -0.265   & $a_{33}$ & -5184.47  & -0.071 \\
$a_{14}$ & -11524.5& -0.081   &  $a_{34}$ & 8260.35   &   0.025   \\
$a_{22}$ & -9856.69   & -0.107  & $a_{44}$ &  -216660.00  & -0.137 \\
\hline
$\beta_1$ & -0.904976 & -27.7 & $\beta_{3}$ & -0.740065 & -10.4  \\
$\beta_2$ & -0.909160&  -28.4  & $\beta_{4}$ & -0.464586   & -4.60  \\
\hline
\end{tabular}
}
\end{table*}

\section{Simulation \label{simulation_sec}}

In this Section we explore the properties of the new robustified ALR-SME and we compare them with the score matching estimator of Scealy and Wood (2022) based on the manifold boundary weight function 
$
\min(u_1, u_2, \ldots u_p, a_c^2),	
$
with various choices of their tuning constant $a_c \in [0,1]$. We consider eight different simulation settings and in each case we simulated $R=1000$ samples and for each sample we calculated multiple different score matching estimates and calculated estimated root mean squared errors (RMSE). The eight different simulation settings are now described. We focus on dimension $p=5$ only.

{\it Simulation 1:} The model is the continuous RPPI model \eqref{uqdensity2} with $\bb_L=\boldsymbol{0}$ and $\beta_5=0$ fixed (not estimated). We set $\beta_1=-0.80$, $\beta_2=-0.85$, $\beta_3=0$, $\beta_4=-0.2$ and $\bA_L$ is equal to the parameter estimates given in Table 3 of Scealy and Wood (2022). This model was the best fitting model for {\it Dataset1} in Scealy and Wood (2022) after they deleted two outliers. We set the sample size to $n=92$ which is consistent with Scealy and Wood (2022). For this model we calculated the new ALR-SME, which is denoted as $c=0$ in Table \ref{tab4}. We also calculated the new robustified ALR-SME with tuning constants set to $c=0.01$ and $c=0.7$. Then we calculated the score matching estimators of Scealy and Wood (2022) with tuning constants set to $a_c=0.01$, $a_c=0.000796$ and $a_c=1$; see columns 5, 6 and 8 in Table \ref{tab4}. Note that $\bbeta$ is not estimated in columns 5, 6 and 8 and is treated as known and set equal to the true $\bbeta$ (Scealy and Wood (2022) treated $\bbeta$ as a tuning constant in their real data application).  The 7th column in Table \ref{tab4} denoted by $a_c$ given $\hat{\bbeta}$ is a hybrid two step estimator. That is, first we calculated the estimate of $\bA_L$ and $\bbeta$ using the robustified ALR-SME with $c=0.7$, then in the second step we updated the $\bA_L$ estimate conditional on the robust $\bbeta$ estimate using the Scealy and Wood (2022) estimator with $a_c=0.000796$. 

{\it Simulation 2:} The model is the multinomial latent variable model with $m_i=2000$ for $i=1,2,\ldots, n$ with $n=92$. The latent variable distribution is set equal to the same RPPI model used in {\it Simulation 1}. We calculated the same score matching estimators as in {\it Simulation 1} but instead of using $\bu_i$ in estimation we plugged in the discrete simulated proportions $\hat{\bu}_i =\bx_i /m_i$. 

{\it Simulation 3:} The same setting as {\it Simulation 1} except we replace $5.4\%$ of the observations with the outlier $\bu_i=(0.4,0.4,0,0,0.2)^{\top}$. 

{\it Simulation 4:} The same setting as {\it Simulation 2} except we replace $5.4\%$ of the observations with the outlier $\hat{\bu}_i=\bx_i/m_i=(0.4,0.4,0,0,0.2)^{\top}$. 

{\it Simulation 5:} The model is the continuous RPPI model \eqref{uqdensity2} with $\bb_L=\boldsymbol{0}$ and $\beta_5=0$ fixed (not estimated), and remaining parameters set equal to the values given in Table \ref{tab3}. For this model we calculated the new ALR-SME which is denoted as $c=0$ in Table \ref{tab6}. We also calculated the new robustified ALR-SME with tuning constants set to $c=0.01$, $c=0.25$, $c=0.5$, $c=0.75$, $c=1$ and $c=1.25$. The sample size is the same as {\it Dataset2} which is $n=94$. 

{\it Simulation 6:} The model is the multinomial latent variable model with $m_i=2000$ for $i=1,2,\ldots, n$ with $n=94$. The latent variable distribution is set equal to the same RPPI model used in {\it Simulation 5}. We calculated the same score matching estimators as in {\it Simulation 5} but instead of using $\bu_i$ in estimation we plugged in the discrete simulated proportions $\hat{\bu}_i =\bx_i /m_i$. 

{\it Simulation 7:} The same setting as {\it Simulation 5} except we replace $5.3\%$ of the observations with the outlier $\bu_i=(0.4,0.3,0.2,0.1,0)^{\top}$.

{\it Simulation 8:} The same setting as {\it Simulation 6} except we replace $5.3\%$ of the observations with the outlier $\hat{\bu}_i=\bx_i/m_i=(0.4,0.3,0.2,0.1,0)^{\top}$.

\begin{table*}[h!]
\centering
{ \footnotesize
\caption{Simulation results {\it Dataset1} RMSE's}
\label{tab4}
\begin{tabular}{|c|c|c|c|c|c|c|c|}
\hline
parameter & $c=0$ & $c=0.01$ & $c=0.7$ & $a_c=0.01$ & $a_c=0.000796$ & $a_c$ given $\hat{\bbeta}$ &  $a_c=1$ \\
\hline
 \multicolumn{8}{c}{{ \it Simulation 1: RPPI model (continuous)}}  \\
\hline
$a_{11}$& 95000  & 88600 & 99000 & 95200 & 81300 & 81900 & 293000  \\
$a_{22}$& 6920 & 6440 & 7720 & 6700 & 5430 & 5460  & 18100  \\
$a_{33}$& 21.9  & 20.5 & 18 & 27.8 & 20.8 & 21.5  & 104 \\
$a_{44}$& 18.7 & 17.7 & 16.6 & 22.6 & 16.2 & 16.6 & 75.6 \\
$a_{12}$& 14800  & 14200 & 22500 & 14800 & 12400 & 12500 & 42900   \\
$a_{13}$& 1240  & 1150 & 974 & 1390 & 1170 &   1190 & 4600 \\
$a_{14}$& 1010 &  963 & 879 &  953 & 807 & 833  & 2690 \\
$a_{23}$& 276 & 262 & 272 & 285 & 242 & 250  &  706 \\
$a_{24}$& 294 & 271 & 238 &  306 & 248 & 251  &  855  \\
$a_{34}$& 15.3 & 14.7 & 14.7 & 16.4 & 12.4 & 13 & 53 \\
$\beta_1$&  0.185  & 0.173 & 0.0685 & - & - & 0.0685 & - \\
$\beta_2$& 0.142 & 0.133 & 0.0581 & - & - &0.0581& -\\
$\beta_3$&  0.264  & 0.255 & 0.245 & - & - & 0.245  & -\\
$\beta_4$&  0.22 & 0.214 &  0.202& - & - & 0.202 & - \\
\hline
 \multicolumn{7}{c}{{ \it Simulation 2: multinomial latent variable model (discrete)}}  \\
\hline
$a_{11}$& 57700  &  55600 & 85700 & 97800 &  97800 & 98200 &  253000  \\
$a_{22}$& 5540  & 5200 & 6240 & 8890 & 8890 &  8910 & 18900 \\
$a_{33}$& 16.5  & 15.7 & 15.1 & 36.7 & 36.7 &  37.1 & 100 \\
$a_{44}$& 16.7 &  15.9 & 15.3 & 36.6 & 36.6 & 36.8  & 84.7 \\
$a_{12}$& 12200  &  11800 & 14300 &  17600 &  17600 & 17700 & 43400   \\
$a_{13}$&  846 & 807 &  844 & 1480 & 1480 & 1490 &  4060  \\
$a_{14}$& 876 & 842 & 803 &  1200 & 1200 &  1210 & 2530 \\
$a_{23}$& 273 & 260 & 241 & 389 & 389 &  392  & 804 \\
$a_{24}$& 248 & 231 &  214 & 417 & 417 &   418  & 882 \\
$a_{34}$& 14.2  & 13.7 &  13.9 & 22.4 & 22.4 &  22.6 & 51.3  \\
$\beta_1$& 0.162 & 0.153 &  0.0631 & - & -& 0.0631  & -\\
$\beta_2$& 0.137  & 0.129 & 0.0543 & - & - & 0.0543 & -\\
$\beta_3$& 0.25 & 0.243 & 0.235 & - & - & 0.235  & -\\
$\beta_4$& 0.214 & 0.209 & 0.197 & - & - & 0.197 & - \\
\hline
\end{tabular}
}
\end{table*}

\begin{table*}[h!]
\centering
{ \footnotesize
\caption{Simulation results {\it Dataset1} RMSE's}
\label{tab5}
\begin{tabular}{|c|c|c|c|c|c|c| c|}
\hline
parameter & $c=0$ & $c=0.01$ & $c=0.7$ & $a_c=0.01$ & $a_c=0.000796$ & $a_c$ given $\hat{\bbeta}$ & $a_c=1$  \\
\hline
 \multicolumn{8}{c}{{\it Simulation 3: RPPI model (continuous) with outliers}}  \\
\hline
$a_{11}$& 3530000  & 93200 & 109000 & 100000 &  85600 & 86500 & 315000 \\
$a_{22}$& 3630000 & 6810 & 7680 & 7000 & 5710 & 5750  & 19600  \\
$a_{33}$&  1090 &  21.4 & 18.8 &  29.2 & 21.7 &  22.4 &  114 \\
$a_{44}$& 2770 & 18.9 &  17.3 & 23.9 & 17.2 &  17.6 & 83.7 \\
$a_{12}$& 3620000  & 15200 &  23300 & 15800 & 13200 &  13400 & 46600   \\
$a_{13}$&  47700 & 1200 & 1020 & 1460 & 1240 & 1250  & 4960  \\
$a_{14}$& 120000 & 1000 & 920 & 1010 &  850 &  880 &  2920 \\
$a_{23}$& 96400 & 277 & 278 & 297 & 255 &  263 & 771 \\
$a_{24}$& 135000  & 289 & 241 & 321 & 263 & 267 & 927 \\
$a_{34}$&  1370 & 15.5 & 15.4 & 17.1 & 13 & 13.5 &  57.3 \\
$\beta_1$& 19.3 & 0.181 & 0.0701 & - & - &  0.0701 & - \\
$\beta_2$& 28.9 & 0.137 & 0.06 & - & - & 0.06 & - \\
$\beta_3$& 17.6  & 0.264 & 0.256 & - & - & 0.256 & -\\
$\beta_4$& 19.2 & 0.224 &  0.211 & - & - &  0.211 &  -\\
\hline
 \multicolumn{8}{c}{{\it Simulation 4: multinomial latent variable model (discrete) with outliers}}  \\
\hline
$a_{11}$& 3050000 & 57200 & 90800 &  105000 & 105000 & 105000 &  281000  \\
$a_{22}$& 3140000 & 5500 & 6680 & 10100 & 10100 & 10100  &  22400 \\
$a_{33}$& 1020 & 16.3 & 15.7 & 41.8 &  41.8 &  42.2 &  114 \\
$a_{44}$& 2360  &  17 & 16.1 & 41.8 & 41.8 & 42.1 & 95.7 \\
$a_{12}$& 3130000 & 12300 & 15600 & 18900 & 18900 & 18900  & 49400 \\
$a_{13}$& 42900 & 825 &  868 & 1630 &  1630 &  1650 &  4590 \\
$a_{14}$& 93500 & 880 & 857 & 1270 & 1270 &  1280  & 2840 \\
$a_{23}$& 85200 & 270 & 251 &  410 & 410 &  413 & 878 \\
$a_{24}$& 116000 & 247 & 225 & 494 & 494 & 495  & 1060 \\
$a_{34}$& 1210  & 14.5 &  14.4 & 25 & 25 & 25.2 &  58.2 \\
$\beta_1$& 16.5  & 0.157& 0.0649 &-  & - & 0.0649  & -\\
$\beta_2$& 27.2  & 0.132 & 0.0555 & - & -& 0.0555  & -\\
$\beta_3$& 16.8 & 0.251 & 0.246 & - & - & 0.246 & -\\
$\beta_4$& 16 & 0.22 & 0.206 & - & - & 0.206  & -\\
\hline
\end{tabular}
}
\end{table*}

\begin{table*}[h!]
\centering
{ \footnotesize
\caption{Simulation results {\it Dataset2} RMSE's}
\label{tab6}
\begin{tabular}{|c|c|c|c|c|c|c|c|}
\hline
parameter & $c=0$ & $c=0.01$ & $c=0.25$ & $c=0.5$ & $c=0.75$ & $c=1$ & $c=1.25$  \\
\hline
 \multicolumn{8}{c}{{\it Simulation 5: RPPI model (continuous)}}  \\
\hline
$a_{11}$& 1950  & 1940 & 1670 & 1490 & 1400 & 1330 & 1840  \\
$a_{22}$& 39900 & 39600 & 34000 & 30300 &  29900 & 34700  & 48100  \\
$a_{33}$& 4790 & 4750 & 3970 & 3740 & 4160 & 5110 & 6080 \\
$a_{44}$& 122000  & 120000 & 97800 & 92700 & 101000 & 128000 & 252000  \\
$a_{12}$& 128000 & 126000 & 107000 & 94600 & 88500 & 104000  & 220000  \\
$a_{13}$& 15300  &  15200 & 13000 & 12000 & 12500 & 14900 & 16000   \\
$a_{14}$& 29200 & 29000 & 24200 &  21500 &  22900 & 26500 & 29600   \\
$a_{23}$& 84900 &  84200 & 70600 & 63200 & 60100 & 61800 &  68400  \\
$a_{24}$& 86200  & 85500 & 74300 & 69400 & 70800 & 77200  & 120000  \\
$a_{34}$& 23300  & 23100 & 20400 & 20300 & 22700 & 28100 & 34900 \\
$\beta_1$& 0.0807  & 0.0796 & 0.0607 & 0.0508 & 0.047 & 0.0462 & 0.0467 \\
$\beta_2$&  0.0869  & 0.0857 & 0.0659 & 0.055 &  0.0496 &  0.0476 & 0.048 \\
$\beta_3$& 0.125  & 0.123 & 0.0931 & 0.0805 & 0.0771 & 0.0786 & 0.0816  \\
$\beta_4$& 0.197 & 0.193 & 0.145 & 0.126 & 0.123 & 0.123 & 0.129 \\
\hline
 \multicolumn{8}{c}{{\it Simulation 6: multinomial latent variable model (discrete)}}  \\
\hline
$a_{11}$& 1840    & 1830   &  1580 & 1410 & 1300 & 1310 & 1330 \\
$a_{22}$& 38300  & 38000  &  32900 & 30800 & 43500 & 49700  & 52300   \\
$a_{33}$& 3950  &3910  & 3290 & 3360 & 3800 & 4500 & 5430 \\
$a_{44}$& 93400  &93500   & 96000 & 97300 & 102000 & 115000  & 143000  \\
$a_{12}$& 176000  & 174000  & 140000 & 116000 & 99900 & 88500   &  87700 \\
$a_{13}$& 12100 & 12000 & 10100 & 9090 &  8660 & 9600  & 11000   \\
$a_{14}$& 24700 & 24500 & 20200 & 18500 & 17100 & 18100  & 19500  \\
$a_{23}$& 72600 & 71800 & 58700 & 50600 & 46100 & 45800   & 46800 \\
$a_{24}$& 80700  & 80000 & 67200 &  59800 &  63700 & 64700  & 69200  \\
$a_{34}$& 18300 & 18100 & 15800 & 16100 & 17400 & 20400  & 24900 \\
$\beta_1$& 0.0634 & 0.0625 & 0.0474 & 0.0395 & 0.036 & 0.0345 & 0.0345  \\
$\beta_2$&   0.0642   &  0.0633 & 0.0476 & 0.0395 & 0.0357 & 0.0349  & 0.0348 \\
$\beta_3$& 0.104  & 0.103 & 0.0789 &  0.0701 & 0.0681 & 0.0675  & 0.0687 \\
$\beta_4$&  0.145  & 0.145 &  0.153 &  0.152 & 0.148 &  0.144 & 0.142 \\
\hline
\end{tabular}
}
\end{table*}

\begin{table*}[h!]
\centering
{ \footnotesize
\caption{Simulation results {\it Dataset2} RMSE's}
\label{tab7}
\begin{tabular}{|c|c|c|c|c|c|c|c|}
\hline
parameter & $c=0$ & $c=0.01$ & $c=0.25$ & $c=0.5$ & $c=0.75$ & $c=1$ & $c=1.25$  \\
\hline
 \multicolumn{8}{c}{{\it Simulation 7: RPPI model (continuous) with outliers}}  \\
\hline
$a_{11}$& 4500000  & 2260 & 1950 & 1740 & 1630 & 1650 & 2490  \\
$a_{22}$& 8.61e+08  & 75200 & 65900 & 60100 & 58400 & 62600  & 82200  \\
$a_{33}$& 3510000 &  4850 & 4050 & 3890 & 4600 & 5250 & 6420 \\
$a_{44}$& 1.24e+08  & 124000 & 101000 & 96300 & 109000 &  132000 & 176000  \\
$a_{12}$& 6.49e+08 & 142000 & 121000 & 107000 & 99700 & 109000  & 256000  \\
$a_{13}$& 58600000  & 16800 & 14200 & 13100 & 13300 & 15300 & 17500   \\
$a_{14}$& 74300000 & 30700 &  25600 & 22900 & 29100 & 32700 & 34800  \\
$a_{23}$& 6.48e+08 & 89400 & 74700 & 66500 & 62100 & 63900 & 70100 \\
$a_{24}$& 6.53e+08  & 94100 & 81400 &  75400 & 76600 &  80200  & 101000  \\
$a_{34}$& 35100000  & 24000 & 21100 & 20900 & 24600 & 28500 & 35800 \\
$\beta_1$& 222  & 0.0826 & 0.0632 & 0.0529 & 0.0487 & 0.048 & 0.0484 \\
$\beta_2$& 222  & 0.0888 & 0.0684 & 0.0573 & 0.0516 & 0.0496 & 0.0498 \\
$\beta_3$& 254  & 0.128 & 0.097 & 0.0841 & 0.0805 & 0.0818 & 0.0833 \\
$\beta_4$& 131 &  0.202 & 0.151 &  0.131 & 0.126 & 0.126 & 0.13 \\
\hline
 \multicolumn{8}{c}{{\it Simulation 8: multinomial latent variable model (discrete) with outliers}}  \\
\hline
$a_{11}$& 2930000  & 2340& 2030 & 1820 & 1500 & 1600 &  1800 \\
$a_{22}$& 3.46e+08  & 62600 & 52800 &  46900 & 56100 & 59600 & 61200  \\
$a_{33}$& 2520000  & 4050 & 3370 &  3340 &  3680 & 4500 &  5300 \\
$a_{44}$& 69800000  & 95000 & 97200 & 98900 & 104000 & 116000 & 143000  \\
$a_{12}$& 2.72e+08 & 182000 &  146000 & 121000 & 104000 & 92300 &   91100 \\
$a_{13}$& 54100000 & 13900 &  11600 & 10200 & 10400 & 10700 & 12300   \\
$a_{14}$& 66200000  & 25400 & 20900 & 18500 & 17700 & 19500 & 20200 \\
$a_{23}$& 2.66e+08 & 79700 & 65200 & 56200 &  51600 & 51200  & 51700 \\
$a_{24}$& 2.76e+08  & 84400 & 70300 & 62300 & 68000 & 70300 &  72700 \\
$a_{34}$&  2.6e+07  & 18600 & 16100 & 16000 & 17200 & 20700 & 25100 \\
$\beta_1$& 175  & 0.0659 & 0.0499 & 0.0415 & 0.0378 & 0.0366 & 0.0363 \\
$\beta_2$& 131  & 0.0663 & 0.05 &  0.0413 & 0.0376 & 0.0367 & 0.0363 \\
$\beta_3$& 186 & 0.108 &  0.0831 & 0.0731 & 0.0701 & 0.0703 & 0.0701 \\
$\beta_4$& 107  & 0.148 &  0.155 & 0.153 & 0.148 & 0.145 & 0.142 \\
\hline
\end{tabular}
}
\end{table*}

We now discuss the simulation results in Tables \ref{tab4}-\ref{tab5}. These models are motivated from {\it Dataset1}. {\it Dataset1} has two components that are highly right skewed concentrated near zero and three components with low concentration; see Figure \ref{figa}. The RMSE's are of a similar order when comparing the first half of Table \ref{tab4} with the corresponding cells in the second half of Table \ref{tab4} and similarly this also occurs within Table \ref{tab5}. This is not surprising because $m_i$ is large compared with $n$ and the approximation $\hat{\bu}_i$ for $\bu_i$ is reasonable. Hence the estimates are insensitive to the large numbers of zeros in $\hat{u}_{i1}$ and $\hat{u}_{i2}$. When comparing Table \ref{tab4} with \ref{tab5} most of the corresponding cells are fairly similar apart from $c=0$ which has huge RMSE's in Table \ref{tab5}. The robustified ALR-SME with $c> 0$ are clearly resistant to the outliers, whereas the unweighted estimator with $c=0$ does not exhibit {good resistance to outliers}. Interestingly, the most efficient estimate of $\bbeta$ is given by $c=0.7$ even when there are no outliers. The efficiency gains for $\beta_1$ and $\beta_2$ are substantial when comparing $c=0$ (no weights) to $c=0.7$. So the message here is that the weighted version of the ALR-SME is valuable for improving efficiency for estimating the components of $\bbeta$ that are negative and close to $-1$. In the continuous case arguably the Scealy and Wood (2022) estimator with $a_c=0.000796$ is the most efficient for estimating $\bA_L$, whereas in the discrete multinomial case the $c=0.7$ estimator is arguably the most efficient for $\bA_L$.

We now consider the simulation results in Tables \ref{tab6}-\ref{tab7}. These models are motivated from {\it Dataset2}. {\it Dataset2} has four components that are highly right skewed concentrated near zero and one component highly concentrated near one; see Figure \ref{figb}. 
The Scealy and Wood (2022) estimators are omitted because their manifold boundary weight functions evaluate to zero, or very close to zero, for most datapoints in most simulated samples.   
The RMSE's are roughly of a similar order when comparing the first half of Table \ref{tab6} with the corresponding cells in the second half of Table \ref{tab6} and similarly this also occurs within Table \ref{tab7}. This is not surprising because $m_i$ is large compared with $n$ and the approximation $\hat{\bu}_i$ for $\bu_i$ is reasonable. Hence the estimates were insensitive to the large numbers of zeros in $\hat{u}_{i1}$, $\hat{u}_{i2}$, $\hat{u}_{i3}$ and $\hat{u}_{i4}$. When comparing Table \ref{tab6} with \ref{tab7} most of the corresponding cells are fairly similar apart from $c=0$ which has huge RMSE's in Table \ref{tab7}. The unweighted ALR-SME is not {resistant to outliers}, whereas the estimators with $c> 0$ are clearly resistant to the outliers. Interestingly, the most efficient estimate of $\bbeta$ is arguably given by $c=1$ or $c=1.25$ even when there are no outliers. Again the message here is that the weighted version of the ALR-SME is valuable for improving efficiency for estimating the components of $\bbeta$ that are negative and close to $-1$. The most efficient estimator for $\bA_L$ is arguably $c=0.5$ or $c=0.75$.

\section{Conclusion}
We proposed a log-ratio score matching estimator that produces consistent estimates for $\bA_L$ and the first $p-1$ elements of $\bbeta$ for the RPPI model and the multinomial model with RPPI latent probability vectors. This estimator was insensitive to the huge number of zeroes often encountered in microbiome data, and even performed well when every datapoint had a component that was zero.
Our new estimator and modelling approach does not require treating zeros as outliers, which is an improvement on the treatment of zeros in the standard Aitchison log-ratio approach based on the logistic normal distribution.
The robustified version of our estimator remained insensitive to zeros, improved resistance to outliers and also improved efficiency over unweighted ALR-SME for well-specified data.
We recommend using our estimators when there are many components, many of which have concentrations at/near zero (i.e. many $\beta_j$, $j=1,2,\ldots, p-1$ are close to $-1$).

\newpage

\appendix

\setcounter{page}{1}

\noindent {\large Supplementary material for the article {\it Robust score matching for compositional data} by Scealy, Hingee, Kent and Wood.}

\section{Appendix}

\subsection{Score matching details \label{details}}

The total number of parameters is $q=p(p-1)/2 + p-1$. Let 
\begin{equation}
\bt(\bu)=(u_1^2,u_2^2,\ldots, u_{p-1}^2,2u_1u_2,2u_1u_3, \ldots, 2u_{(p-2)(p-1)},\log (u_1),\log (u_2),\ldots ,\log (u_{p-1}))^{\top}
\label{tdef}
\end{equation}
denote the sufficient statistics in model (\ref{uqdensity2}) and let 
\begin{equation*}
\bpi=(a_{11},a_{22}, \ldots, a_{(p-1)(p-1)}, a_{12},a_{13},\ldots, a_{(p-2)(p-1)},1+\beta_1, 1+\beta_2\ldots, 1+\beta_{p-1})^{\top}
\end{equation*}
contain the parameters.
Next define the following $q \times (p-1)$ matrix
\begin{equation*}
\bR=\bR(\bu)=
\begin{pmatrix}
\bR_a \\
\bR_b \\
\bR_c
\end{pmatrix},
\end{equation*}
with elements defined as follows. 

Let the index $j=1,2,\ldots, p-1$ represent the columns in $\bR_a$. Each row in $\bR_a$ represents a particular sufficient statistic $u_i^2$ for $i=1,2,\ldots, p-1$ and if $i=j$ then the $i,j$th term in $\bR_a$ is equal to $2u_j^2 (1-u_j)$ or if $i \neq j$ then the $i,j$th term in $\bR_a$ is equal to $-2u_i^2u_j$.

Let the index $k=1,2,\ldots, p-1$ represent the columns in $\bR_b$. Each row in $\bR_b$ represents a particular sufficient statistic $2u_iu_j$ for $i=1,2,\ldots, p-1$ and $j = 1,2, \ldots, p-1$ where $i < j$. So the rows are indexed by the pairs $(i,j)$, that is $(1,2), (1,3), \ldots, ((p-2),(p-1))$. If $k=i$ then the element of $\bR_b$ is equal to $2u_ju_i(1-2u_i)$. If $k=j$ then the element of $\bR_b$ is equal to $2u_iu_j(1-2u_j)$. If $k\neq i$ and $k \neq j$  then the element of $\bR_b$ is equal to $-4u_ku_ju_i$.

Let the index $j=1,2,\ldots, p-1$ represent the columns in $\bR_c$. Each row in $\bR_c$ represents a particular sufficient statistic $\log (u_i)$ for $i=1,2,\ldots, p-1$ and if $i=j$ then the $i,j$th term in $\bR_c$ is equal to $1-u_j$ or if $i \neq j$ then the $i,j$th term in $\bR_c$ is equal to $-u_j$.

Now let $j$ index the columns of $\bR(\bu)$ and $i$ index the rows of $\bR(\bu)$. Define $\bR(\bu)[i,j]$ to be the $i,j$th element of $\bR(\bu)$ and let $\bR(\bu)[,j]$ denote the $j$th column in $\bR(\bu)$. Then 
\begin{equation*}
\bW=\E_0 \left( \sum_{j=1}^{p-1} \bR(\bu)[,j] \left(\bR(\bu)[,j] \right)^{\top}  \right),
\end{equation*}
where the expectation is taken with respect to $f_0 (\bu)$, the true population density 
with respect to Lebesgue measure, $d\Delta^{p-1}$, on $\Delta^{p-1}$.

Next define the following $q \times (p-1)$ matrix
\begin{equation*}
\bS=\bS(\bu)=
\begin{pmatrix}
\bS_a \\
\bS_b \\
\bS_c
\end{pmatrix},
\end{equation*}
with elements defined as follows. 

Let the index $j=1,2,\ldots, p-1$ represent the columns in $\bS_a$. Each row in $\bS_a$ represents a particular sufficient statistic $u_i^2$ for $i=1,2,\ldots, p-1$ and if $i=j$ then the $i,j$th term in $\bS_a$ is equal to $4u_j^2-10u_j^3+6u_j^4$ or if $i \neq j$ then the $i,j$th term in $\bS_a$ is equal to $-2u_i^2u_j+6u_i^2u_j^2$.

Let the index $k=1,2,\ldots, p-1$ represent the columns in $\bS_b$. Each row in $\bS_b$ represents a particular sufficient statistic $2u_iu_j$ for $i=1,2,\ldots, p-1$ and $j = 1,2, \ldots, p-1$ where $i < j$. So the rows are indexed by the pairs $(i,j)$, that is $(1,2), (1,3), \ldots, ((p-2),(p-1))$. If $k=i$ then the element of $\bS_b$ is equal to $2u_ju_i-12u_ju_i^2+12u_i^3u_j$. If $k=j$ then the element of $\bS_b$ is equal to $2u_iu_j-12u_iu_j^2+12u_j^3u_i$. If $k\neq i$ and $k \neq j$  then the element of $\bS_b$ is equal to $-4u_k(1-u_k)u_iu_j+8u_k^2u_iu_j$.

Let the index $j=1,2,\ldots, p-1$ represent the columns in $\bS_c$. Each row in $\bS_c$ represents a particular sufficient statistic $\log (u_i)$ for $i=1,2,\ldots, p-1$. The $i,j$th term in $\bS_c$ is equal to $-u_j(1-u_j)$.

Now let $j$ index the columns of $\bS(\bu)$ and $i$ index the rows of $\bS(\bu)$. Define $\bS(\bu)[i,j]$ to be the $i,j$th element of $\bS(\bu)$ and let $\bS(\bu)[,j]$ denote the $j$th column in $\bS(\bu)$. Then 
\begin{equation*}
\bd=\E_0 \Bigg(  (1+\beta_p)\left(\sum_{j=1}^{p-1} \bR(\bu)[,j] u_j \right)-\sum_{j=1}^{p-1}\bS(\bu)[,j]  \Bigg),
\end{equation*}
where the expectation is taken with respect to $f_0 (\bu)$, the true population density 
with respect to Lebesgue measure, $d\Delta^{p-1}$, on $\Delta^{p-1}$.

The population parameters $\bW$ and $\bd$ can be estimated using sample moments as follows. That is 
\begin{equation}
\hat{\bW}=\frac{1}{n}\sum_{i=1}^n {\bW}_1(\bu_i)
=\frac{1}{n} \sum_{i=1}^n \left( \sum_{j=1}^{p-1} \bR(\bu_i)[,j] \left(\bR(\bu_i)[,j] \right)^{\top}  \right)
\label{Wmom}
\end{equation}
and 
\begin{equation}
\hat{\bd}=\frac{1}{n} \sum_{i=1}^n {\bd}_1(\bu_i)
=\frac{1}{n} \sum_{i=1}^n \Bigg(  (1+\beta_p)\left(\sum_{j=1}^{p-1} \bR(\bu_i)[,j] u_{ij} \right)
-\sum_{j=1}^{p-1}\bS(\bu_i)[,j]  \Bigg).
\label{dmom}
\end{equation}

\subsection{Choices of metric in score matching \label{metric_score}}

The construction of a score matching estimator requires the choice of
a Riemannian metric on the state space. This metric is often taken to
be the Euclidean metric after making a suitable transformation of the
state space. In the context of compositional data, there are three
commonly used transformations of the simplex. Further, for each
transformation there is a choice between using (i) ``embedded
coordinates'', treating the transformed simplex as a
$(p-1)$-dimensional manifold embedded in $\mathbb{R}^p$, or (ii)
``local coordinates'', using the first $(p - 1)$ coordinates to
determine a point on the simplex, noting that the final coordinate is
determined by the first $(p-1)$ coordinates.  Let
$\bu=(u_1,u_2,\ldots, u_p)^{\top}$ be defined on the open simplex,
that is $\bu \in \Delta^{p-1}$ but with the restriction $u_i > 0$, for
$i=1,2,\ldots, p$. Let $z_i=u_i^{1/2}$, $i=1,2,\ldots, p$ denote the
square root transformation and
$v_i=\log{(u_i)}-(1/p)\sum_{j=1}^p \log{(u_j)}$ denote the centered
log-ratio (clr) transformation (Aitchison, 1986).
\begin{itemize}
\item [(LIN)] (i) The ``embedded linear representation'' uses $(u_1,\ldots,u_p)^{\top}$ and views the open simplex as a bounded subset of a hyperplane in $\mathbb{R}^p$. (ii) The ``local linear representation'' uses $(u_1,\ldots,u_{p-1})^{\top}$ and views the open simplex as a bounded open subset in $\mathbb{R}^{p-1}$.
\item [(SRT)] (i) The ``embedded square root'' representation uses $(z_1,\ldots,z_p)^{\top}$ and treats the open simplex as the positive orthant of the unit hypersphere in  $\mathbb{R}^p$. (ii) The ``local square root'' representation uses $(z_1,\ldots,z_{p-1})^{\top}$ and views the open simplex as the positive orthant of the unit ball in $\mathbb{R}^{p-1}$.
\item [(LOG)] (i) The ``embedded log representation'' (or clr representation) uses $(v_1,\ldots,v_p)^{\top}$. This representation fills out a $(p-1)$-dimensional subspace {of $\mathbb{R}^p$}. (ii) The ``local log representation'' (the same as the additive log-ratio or alr representation) uses the $(p -1)$-vector $(v_1 - v_p,\ldots,v_{p-1} -v_p)^{\top}$ which identifies the simplex with all of $\mathbb{R}^{p-1}$.
\end{itemize}

For both of representations (LIN) and (LOG), the mapping between
embedded coordinates and local coordinates involves a linear
transformation. However, this linear transformation is not an
orthogonal transformation, so that the underlying Euclidean metrics
for embedded and local coordinates are different. For the linear and
square root representations, the simplex contains a finite
boundary. Hence it is awkward to construct the score matching
estimator. The reason is that the score matching estimator is based on
Green’s Theorem, and in this case boundary terms appear. Scealy and
Wood (2022) used representation (SRT-i) and resolved the problem of
boundary effects by incorporating an ad hoc weighting function into
the estimation procedure to downweight observations near the boundary.

In contrast this article uses the log representation for which the
boundary has moved to infinity. Hence there are no complications in
the use of Green’s Theorem and the score matching estimator is
straightforward to construct.  This article uses the local coordinates
(LOG-ii) instead of embedded coordinates (LOG-i). There are three
reasons for this choice: (a) using local coordinates simplifies the
description of the estimator’s properties; (b) the RPPI model is only
invariant to the choice of $p-1$ variables (a divisor needs to be
chosen) so using local coordinates based on the first $p-1$ variables is reasonable; and (c) in preliminary comparisons, the score matching
estimator with (LOG-i) exhibited 12\% higher root-mean-square-error
for estimating the Dirichlet component of the model in a situation
with a large proportion (4 out of 5) of the compositional components
highly concentrated close to zero which is relevant in microbiome data
applications.

\subsection{Theorem 2: the influence function \label{influence}}
First define 
\begin{equation}
\bt^{(a)}(\bu)=(u_1^2,u_2^2,\ldots, u_{p-1}^2,2u_1u_2,2u_1u_3, \ldots, 2u_{(p-2)(p-1)},0,0,\ldots ,0)^{\top}.
\label{tdef2}
\end{equation}
Let $F_0$ be the population distribution and let the contaminated version be $F_s=(1-s)F_0+s \delta_{\bz}$, where $\delta_{\bz}$ is a point mass at $\bz \in \Delta^{p-1}$. Define $\bpi_0=\bpi(F_0)$ as the solution to the following population estimating equation
\begin{equation}
\boldsymbol{0}_{q}= \int_{\bu \in \Delta^{p-1}} w(\bu;\bpi_0) \left(  \bW_1(\bu) \bH \bpi_0 -\bd_1 (\bu)   \right)   dF_0(\bu),
\label{pop}
\end{equation}
where $w(\bu;\bpi_0)=\exp{\left( c \bt^{(a)}(\bu)^{\top} \bpi_0   \right)} $ is the weight function {in (\ref{windham2})}, and $\bW_1(\bu)$ and $d_1 (\bu)$ are defined at (\ref{Wmom}) and (\ref{dmom}).

Now define $\bpi_s=\bpi(F_s)$ as the solution to (\ref{pop}), where $F_s$ replaces $F_0$. Hence
\begin{equation*}
\boldsymbol{0}_{q}=\frac{d}{ds} \Big \vert_{s=0} \Bigg\{\int_{\bu \in \Delta^{p-1}} w(\bu;\bpi_s)  \big( \bW_1(\bu) \bH \bpi_s -\bd_1 (\bu)   \big)   dF_s(\bu) \Bigg\}
\end{equation*}
which implies
\begin{equation}
\begin{aligned}
\boldsymbol{0}_{q} = &
 \int_{\bu \in \Delta^{p-1}} \frac{d w(\bu;\bpi_s)}{ds}  \Big \vert_{s=0} \big(  \bW_1(\bu) \bH \bpi_0 -\bd_1 (\bu)  \big)   dF_0(\bu) \\
 & + \int_{\bu \in \Delta^{p-1}} w(\bu;\bpi_0)  \bW_1(\bu) \bH \text{\bf IF}_{\bpi;F_0} (\bz)     dF_0(\bu) \\
 & + w(\bz;\bpi_0) \left(  \bW_1(\bz) \bH \bpi_0 -\bd_1 (\bz)   \right).  
\end{aligned}
\label{estfun}
\end{equation}
Now, 
\begin{equation*}
\frac{d w(\bu;\bpi_s)}{ds}  \Big \vert_{s=0}= w(\bu;\bpi_0) c \left( \bt^{(a)}(\bu)\right)^{\top}  \text{\bf IF}_{\bpi;F_0} (\bz),
\end{equation*}
and plugging this into (\ref{estfun}) we obtain
\begin{equation*}
\begin{aligned}
\boldsymbol{0}_{q} =&
 \int_{\bu \in \Delta^{p-1}}w(\bu;\bpi_0) c \left( \bt^{(a)}(\bu)\right)^{\top}  \text{\bf IF}_{\bpi;F_0} (\bz) 
 \left(  \bW_1(\bu) \bH \bpi_0 -\bd_1 (\bu)   \right)   dF_0(\bu) \\
& + \int_{\bu \in \Delta^{p-1}} w(\bu;\bpi_0)  \bW_1(\bu) \bH \text{\bf IF}_{\bpi;F_0} (\bz)     dF_0(\bu) \\
& + w(\bz;\bpi_0) \left(  \bW_1(\bz) \bH \bpi_0 -\bd_1 (\bz)  \right).  
\end{aligned}
\end{equation*}
Therefore 
\begin{equation}
 \text{\bf IF}_{\bpi;F_0} (\bz)=-\left(\bG(\bpi_0)\right)^{-1}w(\bz;\bpi_0) \left(  \bW_1(\bz) \bH \bpi_0 -\bd_1 (\bz)  \right),
\label{IF}
\end{equation}
where 
\begin{equation}
\begin{aligned}
\bG (\bpi_0) = &
\int_{\bu \in \Delta^{p-1}}w(\bu;\bpi_0) c
\left( \bW_1(\bu) \bH \bpi_0 -\bd_1 (\bu)   \right)
\left( \bt^{(a)}(\bu)\right)^{\top}    dF_0(\bu) \\
& + \int_{\bu \in \Delta^{p-1}} w(\bu;\bpi_0)   \bW_1(\bu) \bH     dF_0(\bu).
\end{aligned}
\label{gdef}
\end{equation}
Note that when $F_0$ is the RPPI distribution with $\beta_j > -1$ and $\beta_j < \infty$ for $j =1,2,\ldots, p-1$ and the eigenvalues of $\bA_L$ are distinct, finite and non-zero then typically $\bG(\bpi_0)$ is finite and the eigenvalues are non-zero. This is a reasonable assumption in general since the elements of $\bG(\bpi_0)$ correspond to different sets of moments of a RPPI variable calculated on a compact space.    Therefore $\bG (\bpi_0)$ is invertible.

\end{document}